\documentclass{article} 
\usepackage{nips15submit_e,times}
\usepackage{hyperref}
\usepackage{url}
\usepackage{tabularx}
\usepackage{booktabs}
\usepackage{graphicx}
\usepackage{subfigure}
\usepackage{fixltx2e}
\usepackage{multirow}
\usepackage{float}

\title{Time Series Analysis of Clickstream Logs from Online Courses}

\author{
Yohan Jo$^*$, Keith Maki$^*$, Gaurav Tomar$^*$\\
(All authors have equal contribution and are listed in alphabetical order by last name.) \\ \\
Language Technologies Institute \\
Carnegie Mellon University \\
\texttt{\{yohanj,kmaki,gtomar\}@cs.cmu.edu}
}

\nipsfinalcopy 

\begin{document}

\maketitle
\begin{abstract}
 Due to the rapidly rising popularity of Massive Open Online Courses (MOOCs), there is a growing demand for scalable automated support technologies for student learning. Transferring traditional educational resources to online contexts has become an increasingly relevant problem in recent years.  For learning science theories to be applicable, educators need a way to identify learning behaviors of students which contribute to learning outcomes, and use them to design and provide personalized intervention support to the students. Click logs are an important source of information about students’ learning behaviors, however current literature has limited understanding of how these behaviors are represented within click logs. In this project, we have exploited the temporal dynamics of student behaviors both to do behavior modeling via graphical modeling approaches and to do performance prediction via recurrent neural network approaches in order to first identify student behaviors and then use them to predict their final outcome in the course. Our experiments showed that the long short-term memory (LSTM) model is capable of learning long-term dependencies in a sequence and outperforms other strong baselines in the prediction task.  Further, these sequential approaches to click log analysis can be successfully imported to other courses when used with results obtained from graphical model behavior modeling.

\end{abstract}

\section{Introduction}
Online courses have become popular over the years. In most courses, clickstream logs are the only data available from all students that gives information about their learning behaviors. While we have abundant learning theories that define educationally meaningful behaviors and learning strategies, we have limited understanding of how their behaviors are represented within click logs. This understanding is important because it allows for the application of traditional educational theories to online settings and helps us identify important study behaviors that lead to positive learning outcomes in online courses. In this project, we model study behaviors from click logs and predict students' final grades based on clicks and behaviors.

Our first task is to model student behaviors using clustering algorithms. Specifically, we focus on how students distribute their time among learning materials in a study session. We cluster click segments using a static clustering method (multinomial mixture model) and a dynamic clustering method (hidden Markov model).

We then predict students' grades using a time series analysis of clicks. There have been studies on predicting learning outcomes in online courses, but to the best of our knowledge, this work is the first to use a time series analysis. Moreover, many of the previous studies focused on information that is not available for all students, e.g., forum activity, whereas this work uses general clickstreams, which are available for all students. We leveraged Long Short Term Memory (LSTM) to learn sequential characteristics relevant to student performance. 

Comparing both sequence-aware and non sequence-aware approaches, we find that approaches which incorporate sequential information outperform those which do not at classifying student performance, and generalize better to other courses.  Additionally, we use these sequence modeling approaches to identify differences between students of different achievement levels, helping to demonstrate how sequence information better distinguishes low graders from high graders.

\section{Related Work}

Existing studies on modeling student behaviors via click logs can be categorized into top-down and bottom-up. Top-down approaches predefine a set of behaviors of interest, such as disengagement and sequential navigation, and corresponding click patterns \cite{Jeske:14,Ramesh:13,Halawa:14}. These approaches provide interpretability, but analyses are focused only on the predefined behaviors and patterns. In contrast, bottom-up approaches aim to find meaningful click patterns from clickstream data and interpret behaviors they mean. For instances, topic modeling approaches~\cite{Wen:14,Coleman:15} treat clicks as words and learn click categories as topics. Using n-grams may give insights onto click sequences that represent the topics. These studies also showed the correlation between the learned patterns and success in the courses. Bottom-up approaches require no predefined set of patterns and learn behaviors that occur frequently, but it may be hard to connect the learned click patterns to meaningful behaviors. 

There are Bayesian network models that can be used for finding meaningful click patterns. Static clustering algorithms, such as Gaussian mixture models or multinomial mixture models, may be used to categorize similar click segments. These models assume that all clusters are independent of each other. Since our data are sequences, dynamic clustering algorithms that consider transitions between clusters may be more appealing. For instances, a hidden Markov model simultaneously clusters similar click segments into states and models transitions between the states. A state transition topic model \cite{Jo:15} extends HMM such that click segments are clustered into topics, and a state is represented as a mixture of the topics. A model for sequential pattern mining \cite{Yang:14} can be used to find frequent click segments. 

There have been a lot of studies that predict students' learning outcomes in online courses. Forum activity, linguistic features in discussions, sentiment, quiz participation, and video interaction have been used as features for dropout prediction \cite{Ramesh:13,Wen:14b,Halawa:14}, certification prediction \cite{Coleman:15}, and performance on quizzes~\cite{Brinton:15}. However, to the best of our knowledge, no work used time series analysis, that is, changes of feature values over time, for prediction.

LSTMs \cite{Hochreiter:97,Graves:12} have become increasingly popular for the task of sequence modeling and time series analysis. They are based on recurrent neural network (RNN) architecture and had been shown to outperform traditional RNNs on numerous temporal processing tasks \cite{Gers:00,Gers:01}. Especially for a sequence modeling task using a traditional RNN, during the gradient back-propagation phase, if the length of the sequences are huge, gradient vanishing or gradient explosion can occur.  We make use of LSTM model which is designed to prevent such situations \cite{Hochreiter:98}. To the best of our knowledge,  this will be the first work to use LSTM to model students' clickstream sequences in an online course.

\section{Methods}

Our goals are twofold. We first want to investigate the general behaviors of students in terms of how they engage in individual learning materials. We then want to examine whether we can predict students' final grades based on the first portion of their click stream, and if we can automatically learn student behaviors which inform the prediction task. In the following sections, we explain the models we used for these two goals.

\subsection{Behavior Modeling}

Different students spend their time differently while interacting with course learning materials, e.g., some students may spend more time on lectures, whereas other students actively engage in forum activity. We hypothesize that the way students distribute their time, which we call \emph{behavior}, throughout a course is closely related to their learning outcomes. More formally we define \emph{session} and \emph{behavior} as follows.

\begin{itemize}
\item Session: a sequence of clicks from one student separated from that student's other click sequences by more than one hour of inactivity.
\item Behavior: a distribution over clicks within a single session.
\end{itemize}

Since behaviors are on an infinite space, we categorize behaviors into a specified number of categories using two clustering algorithms: a multinomial mixture model (MMM) and a hidden Markov model (HMM). 

\paragraph{MMM}
A MMM assumes a fixed set of clusters, each of which has a probability distribution over observations and generates observations from a multinomial distribution. In our task, the clusters are students' states that correspond to individual sessions. Observations are clicks, and the state parameters are the parameters of multinomial distributions that generate clicks. A MMM does not model transitions between successive sessions.

\paragraph{HMM} 
A HMM is a dynamic version of a MMM, considering transitions between successive data points. In this work, a HMM has the same definitions for states and obervations, but each state has two parameters: emission parameter and transition parameter. Emission parameters are equivalent to the state parameters in a MMM, which generates clicks based on multinomial distributions. The transition parameter of each state is a probability distribution of transitions from that state to another state.

\paragraph{}
In Section \ref{sec:experiments}, we will show that students' behaviors learned by these models have a high correlation with their final grades. By comparing the performance by MMM-based behaviors and that by HMM-based behaviors, we will see that taking into account temporal aspects for clustering informs predicting the final grades of students.

\begin{figure}
\centering
\includegraphics[width=.5\linewidth]{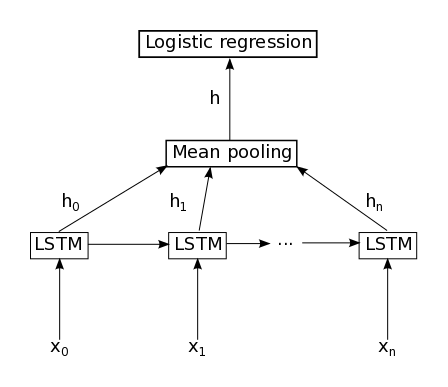}
\caption{The LSTM model used in this work. It is composed of a single LSTM layer followed by mean pooling over time and logistic regression. }
\end{figure}

\subsection{Performance Prediction}

Instructors could benefit from knowing whether a student would eventually perform well or not in the course and then provide personalized support to different supports in form of adaptive interventions. For our work, we used student's grade as final outcome of the course. This section describes the method using student click sequence features in a binary classification task of predicting whether a student achieved a course grade above a threshold. Students who made graded progress towards course completion are assigned an output label of 1, all others are assigned label 0. It should be noted that the data is almost balanced for this classification task, as seen in Table~\ref{tab:data}. If we try to define success of students with a larger threshold than zero, say 40\%, then data will be very skewed (80\% majority class; 20\% minority class).

\paragraph{Baselines} 
We use a baseline \textbf{Support Vector Machine (SVM)} to contextualize the results of our prototype RNN model. 
For this, we use Python's SciKit-Learn module~\cite{scikit-learn}. We ran experiments for SVM using two feature sets: clickstream \textit{length} as the only feature (\textbf{SVM\textsubscript{L}}), and a vector of \textit{counts} of each click type as the features (\textbf{SVM\textsubscript{C}}).

\paragraph{LSTM} Predicting a student's grade from his/her click sequence is a task of sequence modeling and time series analysis, for which RNNs are really popular. To investigate the use of recurrent neural networks in modeling student click sequences, we implemented a LSTM prototype using a Theano-based Deep Learning library, \textit{keras}\footnote{https://github.com/fchollet/keras/}. The implemented model is composed of a single LSTM layer followed by a mean pooling and a logistic regression layer as can be seen in Figure 1. \textit{Hinton's dropout} \cite{Srivastava:14} has been used to prevent over-fitting and sigmoid activation function has been applied to the output. \textit{Adam} optimizer, proposed by Kingma and Lei Ba \cite{Kingma:14} has been used for stochastic optimization and \textit{binary cross entropy} has been used as cost function.

\section{Experiments}\label{sec:experiments}

\subsection{Dataset}
For our experiments, we used click-stream data from two Coursera\footnote{https://www.coursera.org/} courses. 
\begin{itemize}
\item \textbf{Algebra} course which ran from January through April 2013 with 43,361 students. The data from this course was used as training (first 80\%) and validation data (remaining 20\%).
\item \textbf{Pre-Calculus} which ran from January through April 2013 with 51,069 students. The data from this course was used as testing data (100\% data).
\end{itemize}

We preprocessed data to extract click sequences for each student in the course and then performed our experiments for following different set of features.\\
\begin{itemize}
\item{\textbf{Raw clicks}}:
These are the original clicks students made, represented as URLs. 
There are 2648 types of click events in the training/validation datasets, so it is a high dimensional feature set. For instance, this feature set has "view lecture 1" event different from "view lecture 2" event. Note that these features are course-dependent as different courses may have different lectures or assignments, and even have different types of learning activities. Thus, these features are not generalizable and cannot be applied to other courses.
\item{\textbf{Click categories}}:
We further categorized the raw click event types into 46 categories thereby resulting in a low dimensional feature set. For instance, this feature set has "viewed lecture 1" event and "viewed lecture 2" event merged into "viewed a lecture" event. These click categories were decided such that they are consistent across courses and thus generalizable and can be applied to other courses as well.
\item{\textbf{Session states}}:
The click categories were further reduced by dividing the click sequence for a student is divided into sessions (defined earlier) and associating each session with a graphical model state. We decided the number of such states to be 10. These states feature set is almost course-independent and thus the most generalizable among the three feature sets described here.
\end{itemize}

Table~\ref{tab:features} summarizes statistics for the features and labels in the dataset.

\subsection{Behavior Modeling}

This section describes the behaviors learned by MMM and HMM, and interpret them in the context of learning in online courses. We first split each student's click sequence into sessions, and fit all the sequences to MMM and HMM. We use the click categories instead of the raw clicks, because the raw clicks, when fitted to the models, make each state represent learning materials in a similar time period. We empirically chose the number of states to 10. Increasing this number produces redundant behaviors, and decreasing this number reduces the diversity of behaviors.

\begin{figure}
\centering
\subfigure[Behaviors]{
    \includegraphics[width=.36\linewidth]{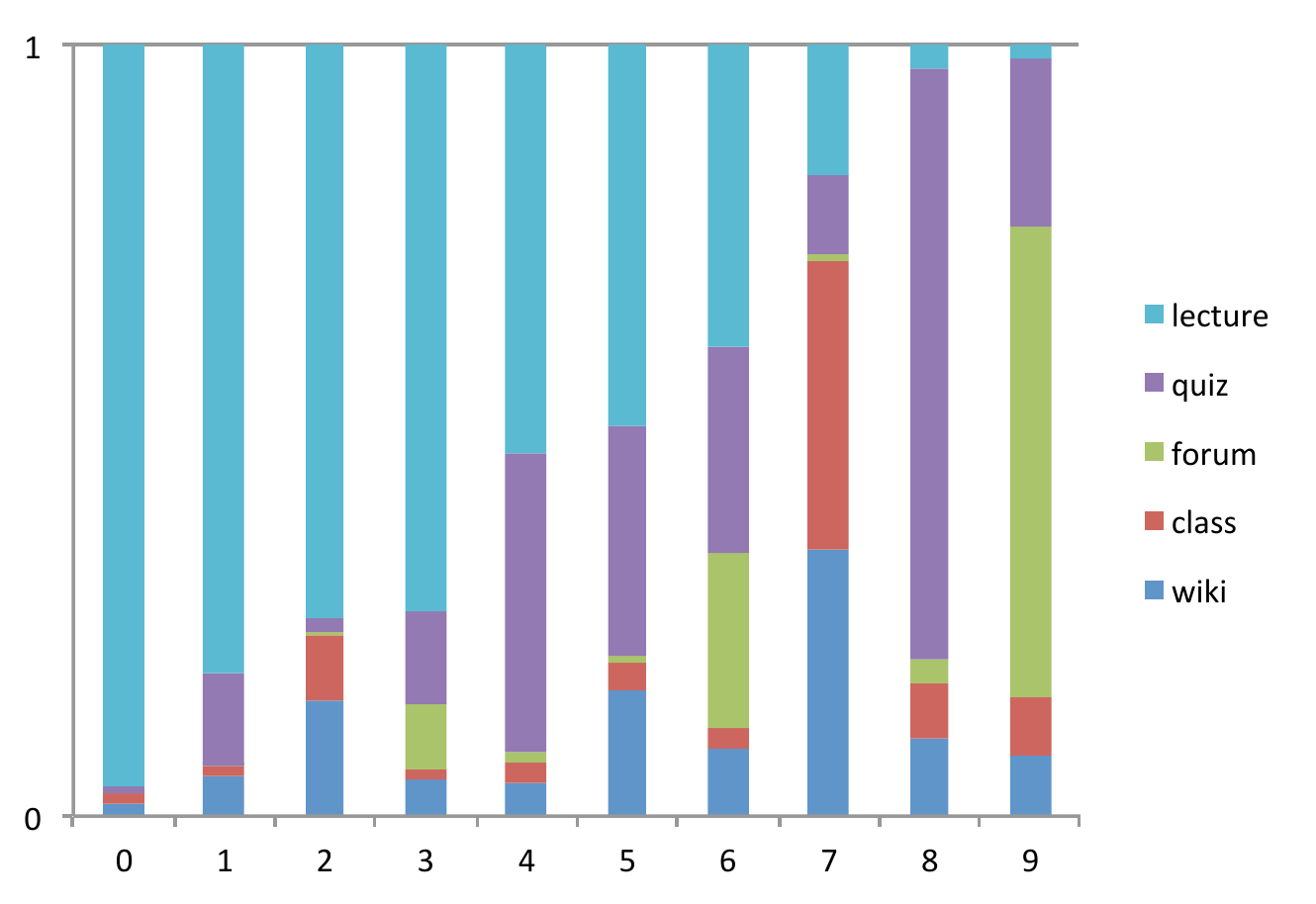}
    \label{fig:behaviors}
}
\subfigure[Behavior Transitions]{
    \includegraphics[width=.5\linewidth]{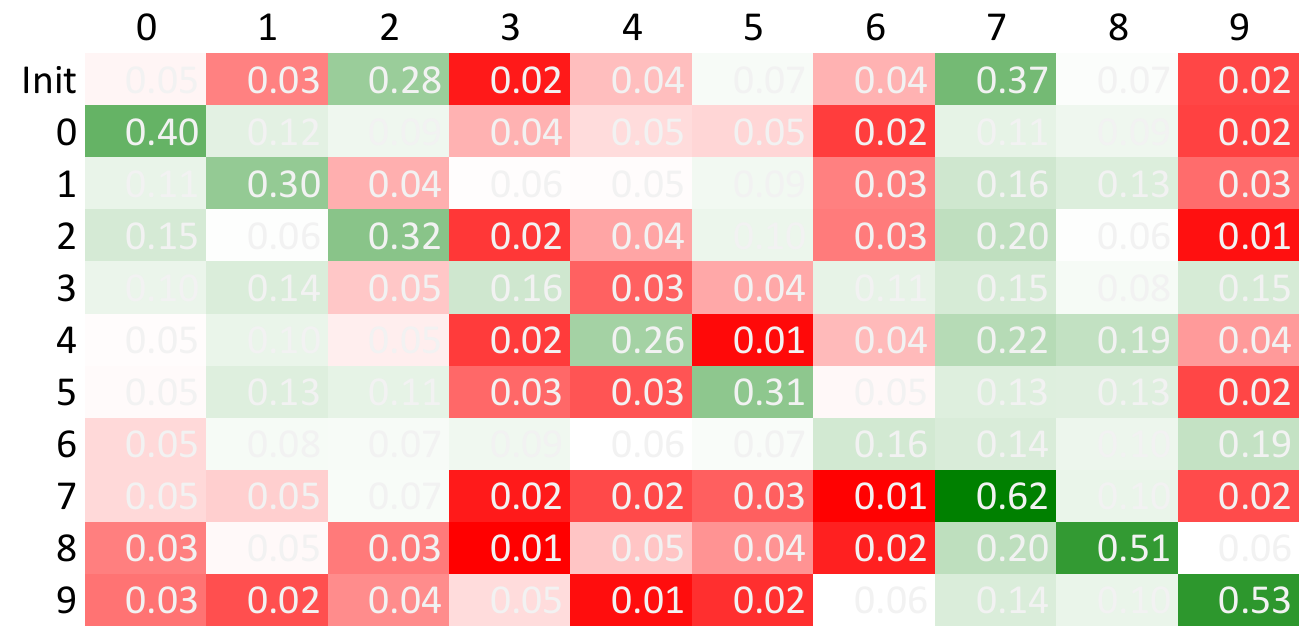}
    \label{fig:transitions}
}
\caption{Behaviors and behavior transitions learned by HMM.}
\end{figure}


The behaviors learned by the MMM and the HMM were very similar. Hence, we demonstrate only the HMM results here. Figure \ref{fig:behaviors} is the learned behaviors. For clear visualization, we combined the click categories into five bigger categories: lecture-, quiz-, forum-, class-, and wiki-related clicks. According to this result, a session is either focused on lectures (states 0-3), composed of lectures and quizzes (states 4 and 5), divided among lectures, quizzes, and forums (state 6), focused on browsing the course (state 7), focused on quizzes (state 8), or focused on forum activity (state 9). 

Figure \ref{fig:transitions} is the learned transitions between the behaviors. The initial probabilities show that many students start their first session with browsing the website (state 8) and additionally watching lectures (state 2). However, students do not tend to take quizzes in their first session. About forum activity, forum-related sessions (states 3, 6, and 9) have high transition probabilities among themselves. For quizzes, a quiz session is very likely to transition to another quiz session (state 8).

\subsection{Feature Analysis}
For qualitative validation of our ability to predict students' grades based on their clicks and behaviors, we examined the features indicative of their final grades.

\begin{table}[t]
    \centering
    \small
    \subfigure[\label{tab:ind_unigrams}Unigrams]{
    \begin{tabularx}{.45\textwidth}{ XX }
        \toprule
        High Graders & Low Graders \\ \hline
        Certificate quiz & Attempt quiz  \\
        Submit quiz & Use quiz late day  \\
        Forum profile & Download lecture \\        
        \bottomrule
    \end{tabularx}
    }
    \subfigure[\label{tab:ind_states}Session States]{
    \begin{tabularx}{.45\textwidth}{ XX }
        \toprule
        High Graders & Low Graders \\ \hline
        State 9 & State 2  \\
        State 3 & State 7 \\
        State 8 & State 0 \\
        \bottomrule
    \end{tabularx}
    }
    \caption{\label{tab:ind}Unigram features indicative of final grades.}
\end{table}
\begin{table}
    \begin{tabularx}{\textwidth}{ XX }
        \toprule
        High Graders & Low Graders \\ \hline
        class/index\_forum/index\_forum/thread & lecture/view\_class/index\_wiki/view  \\
        quiz/feedback\_quiz/index\_quiz/feedback & lecture/view\_lecture/view\_class/index \\
        lecture/view\_lecture/view\_quiz/attempt & wiki/view\_quiz/attempt\_wiki/view \\
        quiz/attempt\_lecture/view\_lecture/view & quiz/feedback\_wiki/view\_quiz/start \\
        quiz/attempt\_quiz/feedback\_forum/index & quiz/start\_quiz/attempt\_wiki/view \\
        \bottomrule
    \end{tabularx}
    \caption{\label{tab:ind_trigrams}Trigram features indicative final grades.}
\end{table}

\paragraph{Unigrams} 
Individual clicks that are frequent among high graders and low graders reveal what learning materials are important to engage in to get a high/low grade in the course. We selected the five most frequent click categories for high graders and low graders, respectively, in each state, and ranked them by popularity. Table \ref{tab:ind_unigrams} shows that students who engage in certificate quizzes and submit quizzes are likely to be high graders. In addition, high graders actively visit other people's profiles in forums, which may indicate their interests in social activity in the course. In contrast, low graders are likely to only attempt quizzes and use late days for quizzes. Interestingly, low graders are more likely to download lecture videos than high graders, which might indicate that they prefer to study whenever they have time, instead of setting aside a time for study.

\paragraph{Trigrams}
Trigrams of clicks (a click segment of length 3) that are frequent among high graders and low graders give insights into the sequences of actions that are related to high grades and low grades. Table \ref{tab:ind_trigrams} shows that high graders engage more in quizzes and forums, whereas low graders engage in lectures and browsing. Some interesting behaviors of high graders are that they start a quiz and consult lectures or they finish a quiz and then go to forums possibly to see other people's opinions on the quiz or to search for relevant information. In contrast, low graders are hesitant to submit a quiz, and they often go back to other pages of the course.

\paragraph{Session States}
States that are frequent among high graders and low graders may inform us of how the distribution of time within a session distinguishes students by grade. Table \ref{tab:ind_states} shows that high graders spend more time on quizzes and forums, whereas low graders primarily view lectures and browse the course. It is not surprising that quizzes are closely related to high grades because grades are based on quiz scores, but it is interesting that forum activity is correlated with higher grades in the course.

\subsection{Performance prediction}
Because many students sign up for MOOCs out of curiosity and do not actively participate in the learning community or even interact substantially with the course material, we consider only students whose clickstreams contain a minimum of 100 clicks . Another motivation behind excluding students with fewer than 100 clicks was to exclude students for whom predicting zero score is trivial. Also, for some students grades are not available, 
so we exclude these students as well.
\\ \\
We partition the students randomly into training (80\%) and test (20\%) partitions, which we use across all experiments.  To understand how well each model can identify struggling students, we learn to predict using only the first part of each clickstream, selected based on the following dimensions:

\begin{itemize}
\item  Number of course days : Days since the course has started. This dimension of evaluation would be useful for an instructor to know at a particular point of time the status of all the students n the course.
\item  Number of student days : Days since a student has joined the course. This dimension of evaluation would be useful for providing out personalized interventions to the student.
\item Number of clicks : Clicks made by a student till now on the course. This dimension gives the more accurate information about the participation about a student.
\item Number of states : The number of states through which student transitioned till now. These states have been fetched from HMM.
\end{itemize}

Each of the four experiments is conducted on the three feature sets: Raw clicks, Click categories, and Session states.
\\ \\
We also examined the transferability of the classifiers trained on the algebra course to the precalculus course. To infer the states of the clickstreams in the new course, we used a maximum likelihood estimation for MMM states and viterbi algorithm for HMM states.

\begin{table*}[t]
    \centering
    \small
    \begin{tabular}{ ccccccccc }
        \toprule
        & \multicolumn{3}{c}{Click Features} & & & \multicolumn{3}{c}{Student Grade Labels} \\
        & Train & Validation & Test & & & Train & Validation & Test \\ \cmidrule{2-4} \cmidrule{7-9}
        \# Clicks & 5,577,375 & 1,311,938 & 6,531,139\protect\footnotemark & & 0 & 42.35\% & 42.77\% & 44.90\% \\
        \# Click Types & 2,378 & 2,021 & 2,623 &&  1 & 57.65\% & 57.23\% & 55.10\% \\ \cmidrule{2-4} \cmidrule{7-9}
        & & & & & \# Instances & 6,749 & 1,688 & 8,984 \\
        \bottomrule
    \end{tabular}
    \caption{Summary of the dataset used for the classification task. The label \textbf{1} denotes "high graders" and \textbf{0} denotes "low graders".\label{tab:data}}
\end{table*}
\footnotetext{Distribution follows Zipf: Linear log-log regression explains 85\% of variance $(p\approx 10^{-225})$.}

\begin{table*}[t]

    \centering
    \small
    \begin{tabular}{ ccccccccc }
        \toprule
        & \multicolumn{3}{c}{Features} \\
        \cmidrule{2-4}
        & Raw clicks & Session states &  Click categories \\ 
        \# Sequences & 8398 & 9155 & 7877 \\
        \# Types &  2648 & 10 & 46 \\ 
        Range &  (101, 56310) & (5,277) & (101, 56310)
 \\ 
        Average &  819 & 23 & 860
  \\ 
        Median &  374 & 13 & 411
 \\ 
        Generalizable? & No & Yes & Most
      \\
        \bottomrule
    \end{tabular}
    \caption{Summary of the feature sets used for the classification task and statistics about corresponding sequences in the data.\label{tab:features}}
\end{table*}

\subsubsection{Results}
Test set accuracies for the performance prediction task are shown in Table~\ref{tab:dev} (Algebra Test set) and Table~\ref{tab:test} (Precalculus Test set).\footnote{Training set accuracies may be found in the appendix (Table~\ref{tab:train}).}  LSTM using raw click features is the clear winner within the Algebra course, significantly outperforming all other baselines for most configurations.  We note, though, that these features are course-dependent, and cannot reliably be used for prediction on other courses (initial experiments using SVM validated this assumption, with a raw click feature baseline performing 5-10 percentage points lower than the MMM state features).  Furthermore, due to the poor performance of click category features even within the Algebra course, it is unlikely that these features will perform well on other courses, even though they may be able to generalize.  However, both MMM and HMM state features show the potential for capturing aspects which distinguish student performance, with LSTM outperforming SVM where sequence information is represented in the features (HMM state features, by a first-order Markov assumption), and failing to beat the baseline when the features assume sessions to be fully independent of one another (MMM state features).  

Table~\ref{tab:test} explores the use of MMM and HMM state features on the precalculus course, to examine whether sequence information improves the generalizability of SVM or LSTM to other courses.  LSTM with HMM features performs surprisingly well, significantly outperforming SVM for a majority of configurations.  
The Multilayer Perceptron (MLP) is a non-RNN option which we trained using the Python Scikit-learn package\footnote{\url{http://scikit-learn.org/dev/modules/generated/sklearn.neural_network.MLPClassifier.html}}. This is a much more difficult baseline, and more investigation is needed to determine whether LSTM is definitively better for generalizability of our approach.  For the Precalculus course data, however, LSTM with HMM features outperforms even this baseline, although the difference is not significant.\footnote{As per a 2-tailed Student T-Test (p$>$0.05).}
Interesting as well is the fact that LSTM significantly outperforms SVM using MMM features in half of the configurations, suggesting that the sequential information encoded in the order of sessions is not completely lost when encoded into MMM features, even though the MMM features themselves make no assumption about ordering.

\begin{table*}[t]
    \centering
    \small
    \begin{tabularx}{\textwidth}{ lX cccccccc }
        \toprule
        && \multicolumn{4}{ c }{Click Features} & \multicolumn{4}{ c }{State Features} \\ \cmidrule{4-9}
        && \multicolumn{2}{ c }{Raw Clicks} & \multicolumn{2}{ c }{Click Categories}  & \multicolumn{2}{ c }{MMM} & \multicolumn{2}{ c }{HMM}  \\ \cmidrule{3-7} \cmidrule{8-10}
        &&  LSTM  & SVM & LSTM  & SVM & LSTM  & SVM & LSTM  & SVM  \\ \midrule
        \multirow{4}{*}{Course Days} & 7 & \textbf{71.49} & 70.71 & 70.30 & 68.65 & 69.20 & 69.95 & \textbf{71.87} & 71.22\\
        & 18 & \textbf{85.65} & 83.69 & 70.37 & 73.03 & 78.48 & 80.67 & \textbf{81.16} & 79.19\\ 
        & 35 & \textbf{92.80*\protect\footnotemark} & 89.46 & 74.68 & 74.56 & 85.20 & \textbf{87.58} & 86.73 & 84.82\\ 
        & All & \textbf{94.79*} & 90.77 & 73.60 & 76.02 & 87.00 & \textbf{91.23*} & 88.26 & 86.78\\ 
        \midrule
        \multirow{4}{*}{Student Days} & 7 & 80.89 & \textbf{81.01} & 75.82 & 75.19 & 75.86 & \textbf{77.43} & 71.87 & 71.22\\ 
        & 18 & \textbf{93.57*} & 88.69 & 80.77 & 76.08 & 84.00 & \textbf{86.17} & 81.16 & 79.19\\ 
        & 35 & \textbf{94.94*} & 90.30 & 74.43 & 75.32 & 86.89 & \textbf{89.88*} & 86.73 & 84.82\\ 
        & All & \textbf{94.79*} & 90.77 & 73.60 & 76.02 & 87.00 & \textbf{91.23*} & 88.26 & 86.78\\ 
        \midrule
        \multirow{4}{*}{\#clicks} & 100 & \textbf{92.24*} & 71.85 & 75.82 & 69.10 & & & &\\
        & 1000 & \textbf{94.33*} & 90.77 & 76.40 & 75.06 & & & &\\ 
        & 1959 & \textbf{94.79*} & 90.89 & 77.60 & 76.02 & & & &\\ 
        & All & \textbf{96.15*} & 91.01 & 78.74 & 76.59 & & & &\\ 
        \midrule
        \multirow{4}{*}{\#states} & 10 & & & & & 80.56 & 84.38 & 83.66 & \textbf{85.09}\\ 
        & 25 & & & & & 86.67 & 86.46 & \textbf{87.88} & 87.55\\ 
        & 50 & & & & & 86.78 & 86.40 & \textbf{88.46} & 86.78\\ 
        & All & & & & & 87.66 & 86.07 & \textbf{88.64} & 86.78\\ 
        \bottomrule
    \end{tabularx}
    \caption{Classification performance on the algebra course.\label{tab:dev}}
\end{table*}

\footnotetext{Statistical significance as determined by a 2-tailed Student T-Test (p$\le$0.05).}

\begin{table*}[t]
    \centering
    \small
    \begin{tabularx}{0.8\textwidth}{ lX cccccc }
        \toprule
        && & \multicolumn{4}{ c }{State Features} \\ \cmidrule{4-7}
        && \multicolumn{3}{ c }{MMM} & \multicolumn{3}{ c }{HMM}  \\ \cmidrule{3-8}
        &&  LSTM & SVM & MLP & LSTM & SVM & MLP  \\ \midrule
        \multirow{4}{*}{Course Days} & 7 & 72.26 & 73.38* & & \textbf{74.13} & 73.72\\ 
        & 18 & 79.99 & 79.86 & & \textbf{81.79} & 81.06 &\\ 
        & 35 & 85.21* & 84.19 & & \textbf{86.71*} & 85.50 &\\ 
        & All & 80.43 & 85.19* & & \textbf{89.46*} & 87.10 &\\ 
        \midrule
        \multirow{4}{*}{Student Days} & 7 & 77.55 & 78.17 & & \textbf{78.94} & 78.67\\ 
        & 18 & 84.35* & 83.34 & & \textbf{85.52*} & 84.26 & \\ 
        & 35 & 86.99* & 85.20 & & \textbf{88.34*} & 86.98 & \\ 
        & All & 86.35* & 85.19 & & \textbf{89.46*} & 87.10 & \\ 
        \midrule
        \multirow{4}{*}{\#states} & 10 & 80.72 & 83.63* & 84.27*\protect\footnotemark & 83.37 & \textbf{85.85*} & 85.68*\\ 
        & 25 & 87.30 & 86.05 & 87.77* & \textbf{88.44} & 88.15 & 88.27\\ 
        & 50 & 87.77* & 85.38 & 87.48* & \textbf{89.15*} & 87.09 & 88.55*\\ 
        & All & 87.75* & 85.19 & 87.69* &\textbf{89.46*} & 87.10 & 88.39*\\ 
        \bottomrule
    \end{tabularx}
    \caption{Classification performance on the precalculus course.\label{tab:test}}
\end{table*}
\footnotetext{Indicates significant improvement over the lowest performing classifier.}

\section{Discussion \& Future Work}
The work presented compares LSTM to simple baselines to demonstrate the strength of sequential feature information in modeling and capturing characteristics of sequential inputs such as clickstream data.  However, we acknowledge the limitations of this work with respect to the comparisons which were made.

First, the two Coursera courses used came from the same institution (UC-Irvine), and overlap substantially in terms of content, even sharing some of the same video lectures.  Additionally, these courses were held at the same time, which controls for many platform-dependent aspects specific to the version of Coursera which was live at that time.  Future work should extend the comparisons we present here to other courses which are not held at same time, on the same platform, by the same instructors, or even on similar topics.  A truly generalized theory of student learning would apply across all these domains, and it would be interesting to see the extent to which LSTMs or other methods could generalize in this way.

We conducted some initial experiments which showed that LSTM outperformed a simple neural network baseline in generalizing to other course data, but the results were not conclusive enough to claim statistical significance.  Future work should investigate whether the success of LSTMs is due to the qualities of neural networks, or due to the sequential applicability of its framework.

The behavior modeling techniques explored so far consider only states which give equal weight to each click emitted by a state. Future work could consider state emissions which take into consideration the length of time a student spends on each click, acknowledging the intuition that students who take more or less time to complete the same activities may have differing course performance.

The MMM features gives a good sequence-agnostic comparison for HMM, but initial experiments using GMM features suggest this or another sequence-agnostic model may have a better fit to the data than MMM features.  Future work should explore the use of GMM features or Gaussian HMM emissions to examine whether these features could surpass those explored here.

\section{Conclusion}

In this work, we demonstrated modeling students' behaviors in online courses via click logs, and predicted students' final grades on the basis of their clicks and behaviors. The experiments revealed that students with a high grade actively engage in forums and quizzes, whereas low-grade students tend to watch or download lectures and attempt quizzes without submission. These course-independent behaviors turned out to achieve a high accuracy in predicting final grades and to generalize well to other courses. Although raw clicks were the most informative of final grades, they cannot be used in other courses because they are incorporated with course-specific information. On the other hand, click categories are course-independent, but they predict final grades poorly. Another contribution of this work is that we increased prediction accuracy by using the temporal information of clicks. This arguably suggests that the temporal dynamics of clicks convey information about final grades that cannot be obtained from individual clicks.

\pagebreak

\section*{Appendix}

\begin{table*}[h!]
    \centering
    \small
    \begin{tabularx}{\textwidth}{ lX cccccccc }
        \toprule
        && \multicolumn{4}{ c }{Click Features} & \multicolumn{4}{ c }{State Features} \\ \cmidrule{4-9}
        && \multicolumn{2}{ c }{Raw Clicks} & \multicolumn{2}{ c }{Click Categories}  & \multicolumn{2}{ c }{MMM} & \multicolumn{2}{ c }{HMM}  \\ \cmidrule{3-7} \cmidrule{8-10}
        &&  LSTM  & SVM & LSTM  & SVM & LSTM  & SVM & LSTM  & SVM  \\ \midrule
        \multirow{4}{*}{Course Days} & 7 & 71.78 & 75.50 & 69.21 & 81.37 & 68.27 & 69.95 & 71.35 & 71.30\\ 
        & 18 & 85.58 & 88.03 & 70.00 & 91.99 & 79.22 & 80.67  & 81.74 & 81.27\\ 
        & 35 & 92.57 & 94.64 & 72.31 & 97.27 & 84.74 & 87.58  & 87.18 & 88.04\\ 
        & All & 95.00 & 97.89 & 75.00 & 99.75 & 86.97 & 91.23 & 89.01 & 91.30\\ 
        \midrule
        \multirow{4}{*}{Student Days} & 7 & 80.90 & 86.41 & 72.69 & 95.43 & 75.44 & 77.43 & 78.56 & 78.45\\ 
        & 18 & 95.07 & 94.83 & 81.62 & 98.54 & 83.25 & 86.17  & 86.17 & 86.66\\ 
        & 35 & 95.01 & 97.23 & 72.88 & 99.37 & 86.77 & 89.88  & 88.07 & 90.03\\ 
        & All & 95.00 & 97.89 & 75.00 & 97.75 & 86.99 & 91.23 & 89.01 & 91.29\\ 
        \midrule
        \multirow{4}{*}{\#clicks} & 100 & 92.74 & 73.67 & 75.48 & 79.59 & & & &\\ 
        & 1000 & 94.48 & 97.89 & 76.40 & 99.76 & & & &\\ 
        & 1959 & 95.00 & 97.84 & 77.71 & 99.75 & & & &\\ 
        & All & 94.75 & 97.84 & 79.00 & 99.73 & & & &\\ 
        \midrule
        \multirow{4}{*}{\#states} & 10 & & & & & 81.29 & 85.53 & 83.72 & 85.84\\ 
        & 25 & & & & & 86.77 & 90.31 & 87.90 & 90.70\\ 
        & 50 & & & & & 86.93 & 91.19 & 89.14 & 91.30\\ 
        & All & & & & & 87.00 & 91.23 & 87.71 & 91.30\\ 
        \bottomrule
    \end{tabularx}
    \caption{Training set classification task performance of the baselines and LSTM.\label{tab:train}}
\end{table*}

\begin{figure}[h!]
\centering
\subfigure[Course days vs. Accuracy]{
    \includegraphics[width=.99\linewidth]{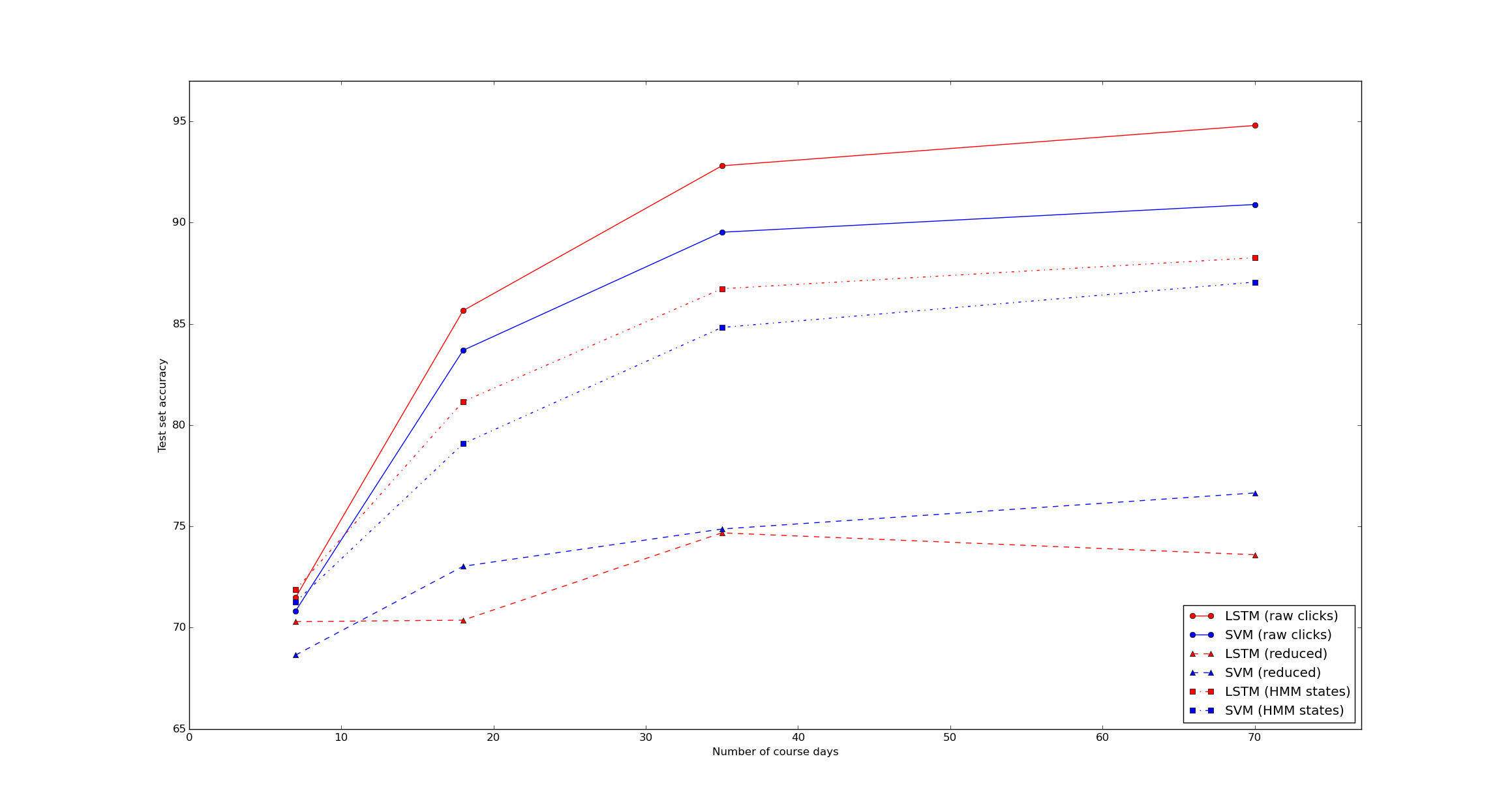}
    \label{fig:Course days vs. Accuracy}
}
\subfigure[Student Days vs. Accuracy]{
    \includegraphics[width=.99\linewidth]{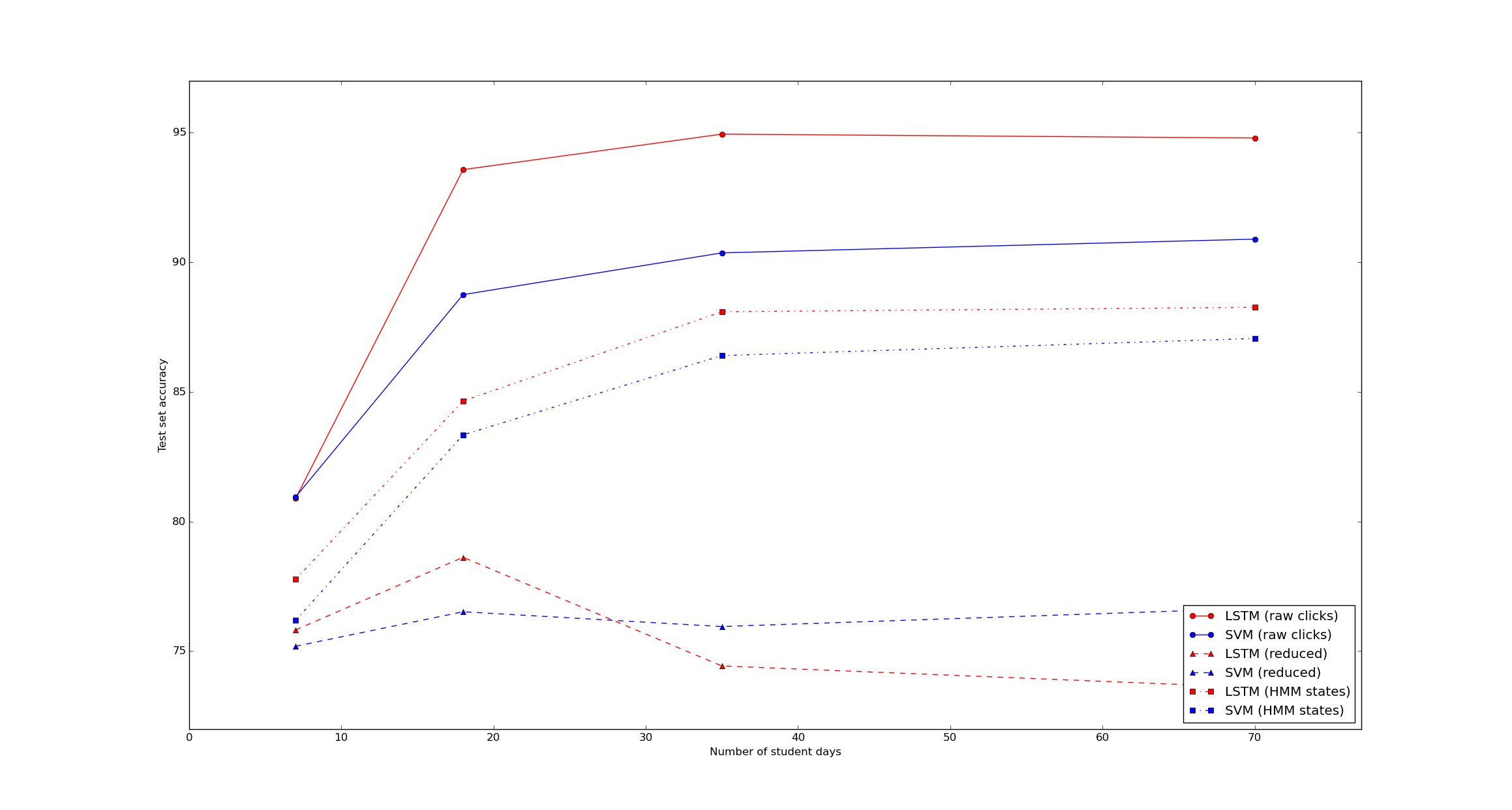}
    \label{fig:Student Days vs. Accuracy}
}

\caption{Grade prediction performance across time. Horizontal axis is number of days (course days/student days) and vertical axis is the accuracy.}
\end{figure}

\begin{figure}[h!]
\centering

\subfigure[No. of clicks vs. Accuracy]{
    \includegraphics[width=.99\linewidth]{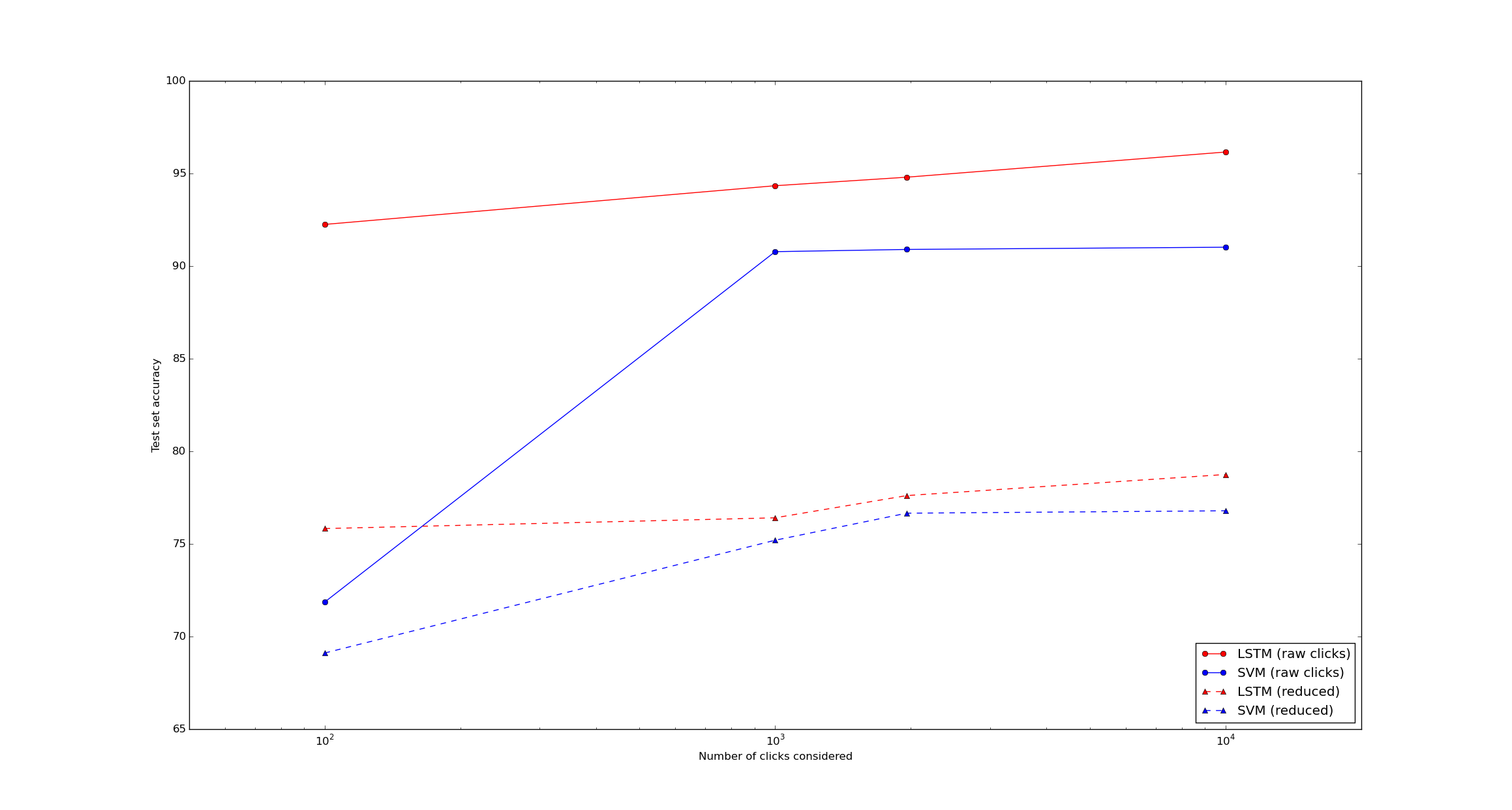}
    \label{fig:No. of clicks vs. Accuracy}
}
\subfigure[No. of states vs. Accuracy]{
    \includegraphics[width=.99\linewidth]{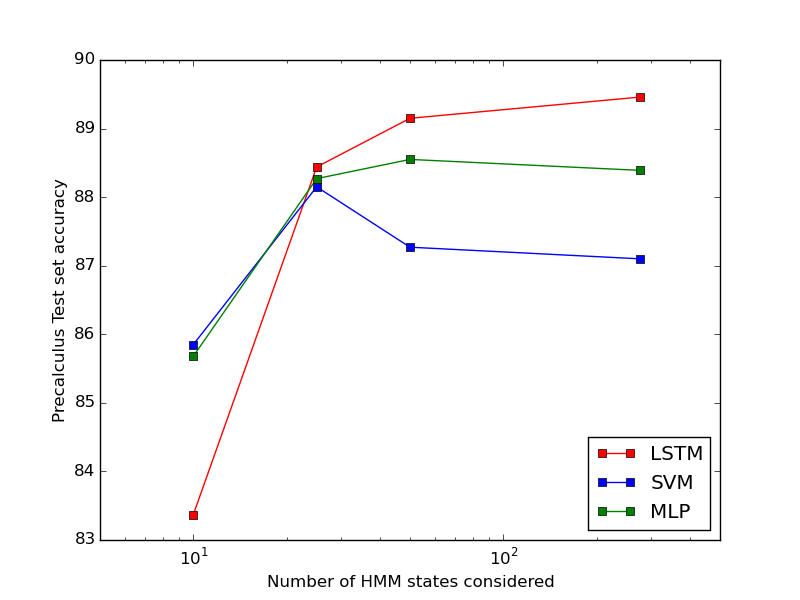}
    \label{fig:No. of states vs. Accuracy}
}
\caption{Grade prediction performance across clicks or states. Horizontal axis is number of clicks or states and vertical axis is the accuracy. For more details, see Table~\ref{tab:test} and related discussion.}
\end{figure}

\end{document}